\documentclass[slac_one]{revtex4}
\usepackage{graphicx}
\usepackage{fancyhdr}
\begin{document}

\title{Searches for New Physics at H1}

%

\author{D.~M.~South\footnote{on behalf of the H1 Collaboration.}}
\affiliation{Technische Universit\"at Dortmund, Experimentelle Physik V, 44221 Dortmund, Germany}

\begin{abstract}
Recent results of searches for leptoquarks, lepton flavour violating leptoquarks
and excited fermions (electrons, neutrinos and quarks) with the
H1 experiment at HERA are presented, which use up to the full $e^{\pm}p$ H1 data set.
No evidence for the direct or indirect production of such particles is
found.
The results are interpreted in terms of limits on the Yukawa coupling
of leptoquarks and lepton flavour violating processes and for excited
fermions on the ratio of the coupling parameter to the compositeness scale
$f/\Lambda$, mainly in the framework of gauge mediated interactions.
The derived limits extend the excluded regions to higher masses than those
reported in previous results.
\end{abstract}

\maketitle

\thispagestyle{fancy}

\section{INTRODUCTION}

The HERA $e^{\pm}p$ collider, located in Hamburg, Germany, was in
operation in the years $1992$--$2007$.
Protons with an energy up to $920$~GeV were brought into collision with
electrons or positrons of energy $27.6$~GeV at two experiments,
H1 and ZEUS, each of which collected a data sample with a total integrated 
luminosity of about $0.5$~fb$^{-1}$.
The deep inelastic scattering (DIS) interactions produced at HERA, at a centre
of mass energy $\sqrt{s}$ of up to $319$~GeV, provided an ideal environment to
study rare processes, set constraints on the Standard Model (SM) and search for
new particles and physics beyond the Standard Model (BSM).
Searches performed by the H1 Collaboration for leptoquarks, lepton flavour
violating leptoquarks and excited fermions are presented in the following.
Analyses of other rare processes at HERA are described elsewhere in
these proceedings.

\section{SEARCHES FOR LEPTOQUARKS}

The HERA data provide the unique possibility to search for the resonant
production of new particles that couple directly to a lepton and a parton.
Leptoquarks (LQs), which appear naturally in various unifying BSM theories,
are such an example.
At HERA, LQs may be resonantly produced in the $s$-channel up to the centre of
mass energy or virtually exchanged in the $u$-channel between the initial state
lepton and a quark coming from the proton.
Beyond the centre of mass energy the production mechanism proceeds via contact
interactions.

A search for LQ production is performed by H1 in the $14$ LQ framework of
the Buchm\"{u}ller, R\"{u}ckl and Wyler (BRW) model~\cite{brw}, using the full
H1 HERA data set.
LQs are classified in terms of fermion number $F=|3B+L|$, where $B$ and $L$ are the
baryon and lepton number respectively.
Due to the more favourable density of quarks with respect to
anti--quarks at high $x$, the $e^{-}p$ data are mostly sensitive to LQs with
fermion number $F=2$, whereas the $e^{+}p$ data are more sensitive to $F=0$ LQs.
H1 results based on HERA~I data are presented in~\cite{h1lq05}.

\begin{figure*}[t]
\centering
\includegraphics[width=0.45\columnwidth]{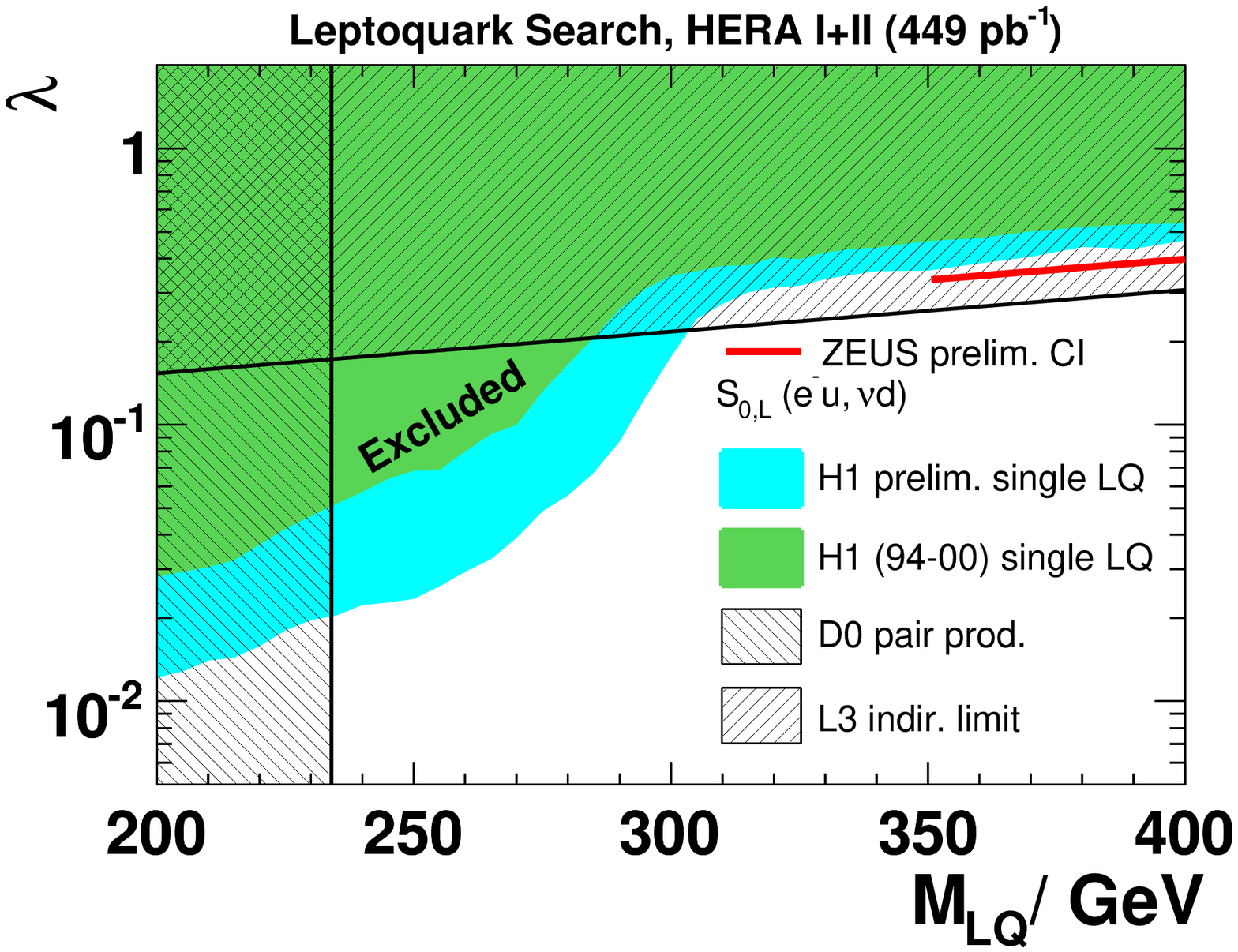}
\includegraphics[width=0.45\columnwidth]{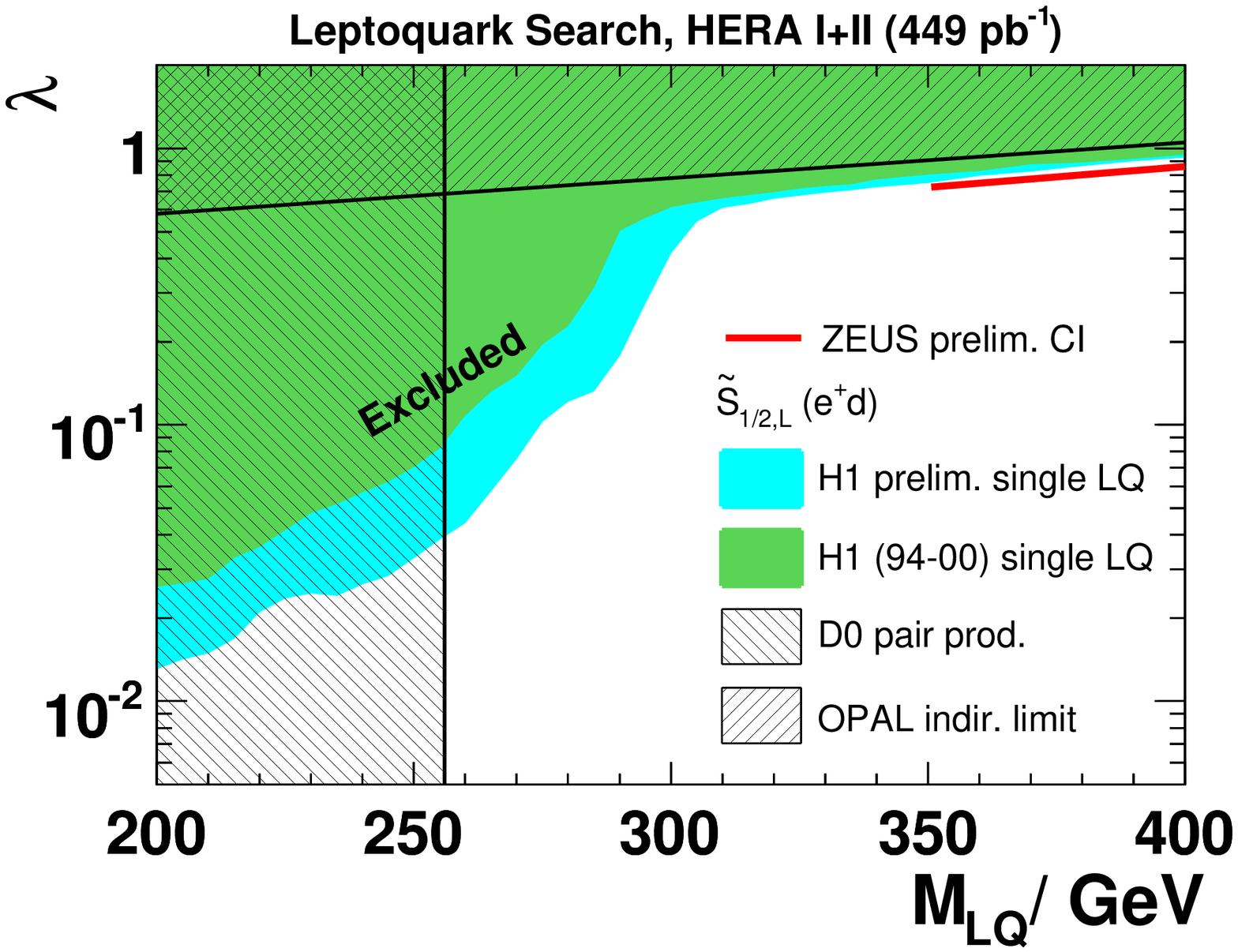}
\caption{H1 exclusion limits at $95\%$~C.L. on the Yukawa coupling
    $\lambda$ as a function of the LQ mass for the $S_{0,L}$ (left) and
    $\tilde{S}_{1/2,L}$ (right) leptoquark in the BRW model.
    The improvement with respect to previous HERA~I H1 results~\cite{h1lq05} is also visible.}
\label{fig:lq}
\end{figure*}

In the search for first generation LQs, the processes $ep \rightarrow LQ \rightarrow eq$
and $ep \rightarrow LQ \rightarrow \nu q$ are considered.
These LQ decays lead to final states similar to those of neutral current (NC) and charged
current (CC) DIS interactions, which constitute an irreducible SM background.
No signal is observed in the DIS mass spectra and constraints on LQ production are set,
which extend beyond the previously excluded domains.
For a Yukawa coupling $\lambda$ of electromagnetic strength, LQ masses below
$291$--$330$~GeV are ruled out by H1, depending on the LQ type~\cite{h1lq}.
Figure~\ref{fig:lq} presents the constraints on $\lambda$ for the
$S_{0,L}$ and $\tilde{S}_{1/2,L}$ LQs obtained by H1 as a function of LQ mass, $M_{LQ}$.
Limits from LEP~\cite{leplq} are also shown, derived from indirect constraints
on the process $e^{+}e^{-}\rightarrow q\overline{q}$, as well as limits from the
Tevatron~\cite{d0lq}, where LQs would be pair--produced via the strong interaction
resulting in a production rate that is independent of $\lambda$.
As mentioned above, contact interactions can also be used to describe the
effects of virtual LQ production or exchange at HERA in the limit of large
LQ mass $M_{LQ} \gg \sqrt{s}$ and indirect LQ limits from the ZEUS contact
interaction analysis~\cite{zeusCI} are also illustrated in figure~\ref{fig:lq}.

\begin{figure*}[h]
\centering
\includegraphics[width=0.42\columnwidth]{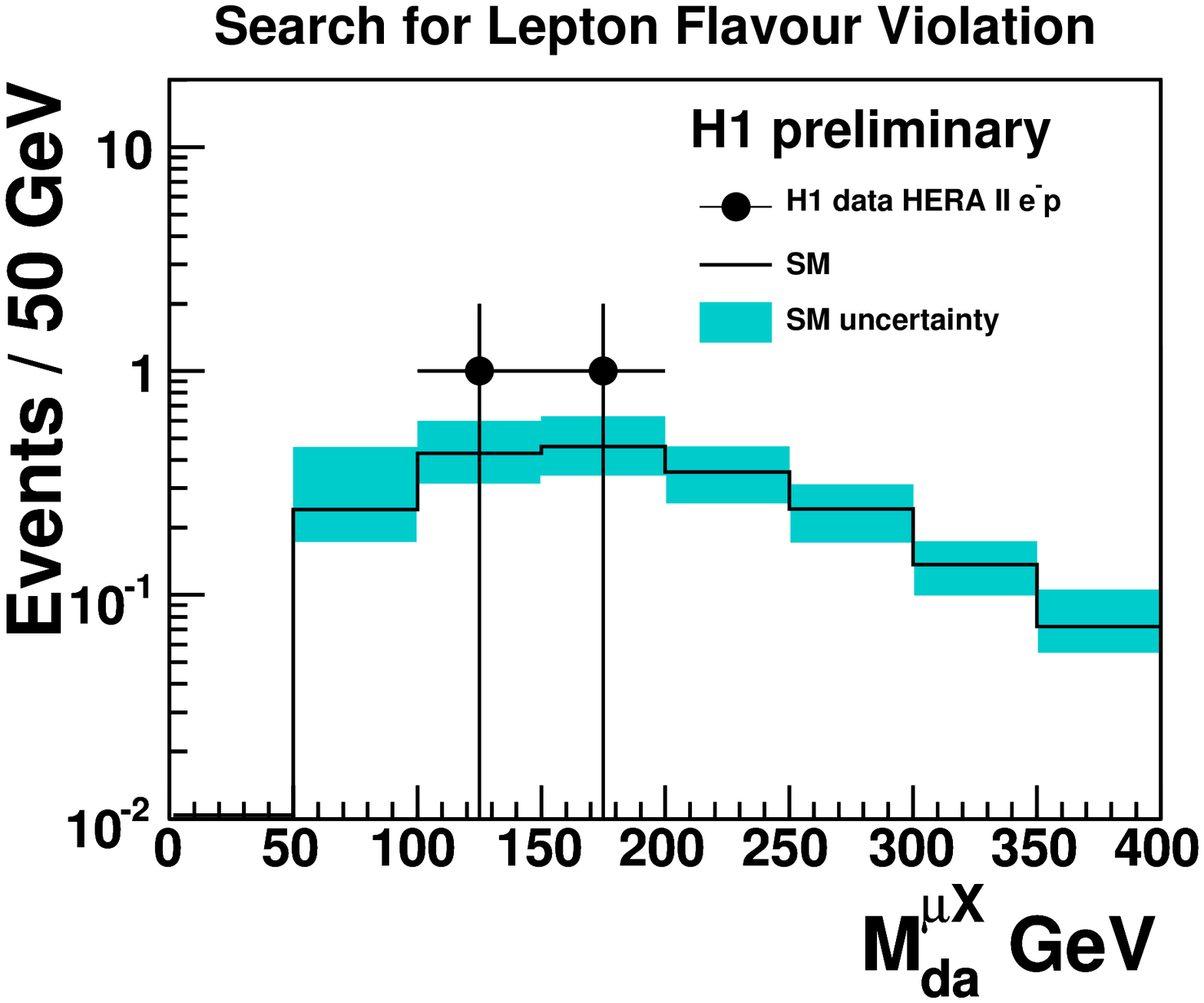}
\includegraphics[width=0.40\columnwidth]{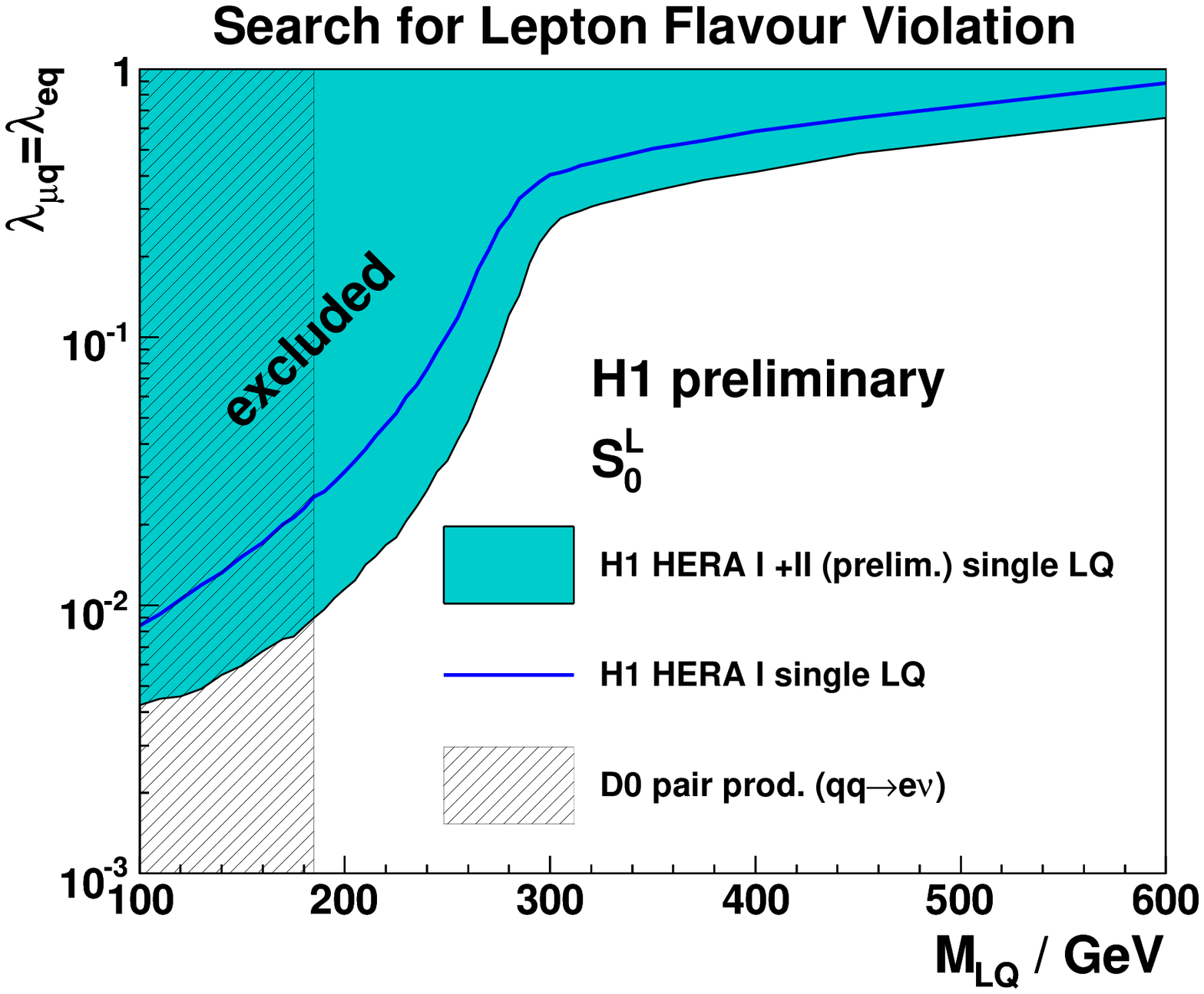}
\caption{Left: Leptoquark mass spectrum after the final selection. The histogram shows the SM expectation,
  and the shaded band the SM uncertainty. Right: Exclusion limits at $95\%$~C.L. on the coupling
  $\lambda_{\mu q} = \lambda_{eq}$ as a function of LQ mass for the $S_{0,L}$ leptoquark in the BRW model.
  The improvement with respect to previous HERA~I H1 results~\cite{h1lfv07} is also visible.}
\label{fig:lfv}
\end{figure*}

A search for lepton flavour violating (LFV) processes in $ep$ collisions
mediated by LQs is also performed by H1.
At HERA, the LFV processes $eq \rightarrow LQ \rightarrow \mu q$ and
$eq \rightarrow LQ \rightarrow \tau q$ may lead to final states with a
muon or tau lepton, together with a hadronic jet.
H1 results based on HERA~I data are presented in~\cite{h1lfv07}, where
searches for both second and third generation LQs are included.
A new search for second generation $F=2$ LQs using the complete H1 HERA~II
$e^{-}p$ data is presented here~\cite{h1lfv}.
Events are selected with a high $P_{T}$ muon and large missing calorimetric $P_{T}$.
The resulting LQ mass spectrum is shown in figure~\ref{fig:lfv}~(left).
Two candidate data events are observed in agreement with the SM prediction,
which, in contrast to the first generation, is negligible.
In the absence of a LQ signal, limits are derived for the $7$ $F=2$ LQ types
as a function of $M_{LQ}$.
For Yukawa couplings of electromagnetic strength, LFV LQs coupling to a muon--quark pair
are ruled out for LQ masses up to $433$~GeV.
Figure~\ref{fig:lfv}~(right) presents the constraints for the coupling
$\lambda_{\mu q} = \lambda_{eq}$ for the $S_{0,L}$ LQ obtained by H1 as a
function of $M_{LQ}$.
A direct limit obtained at the TeVatron from leptoquark pair production is
also shown for comparison~\cite{d0lfv}.

\section{SEARCHES FOR EXCITED FERMIONS}

The existence of excited states of leptons and quarks is a natural
consequence of models assuming composite fermions, and their discovery
would provide convincing evidence of a new scale of matter.
The production and decay of such particles is described in
gauge--mediated (GM) and contact--interaction (CI) models.

\begin{figure*}[h]
\centering
\includegraphics[width=0.41\columnwidth]{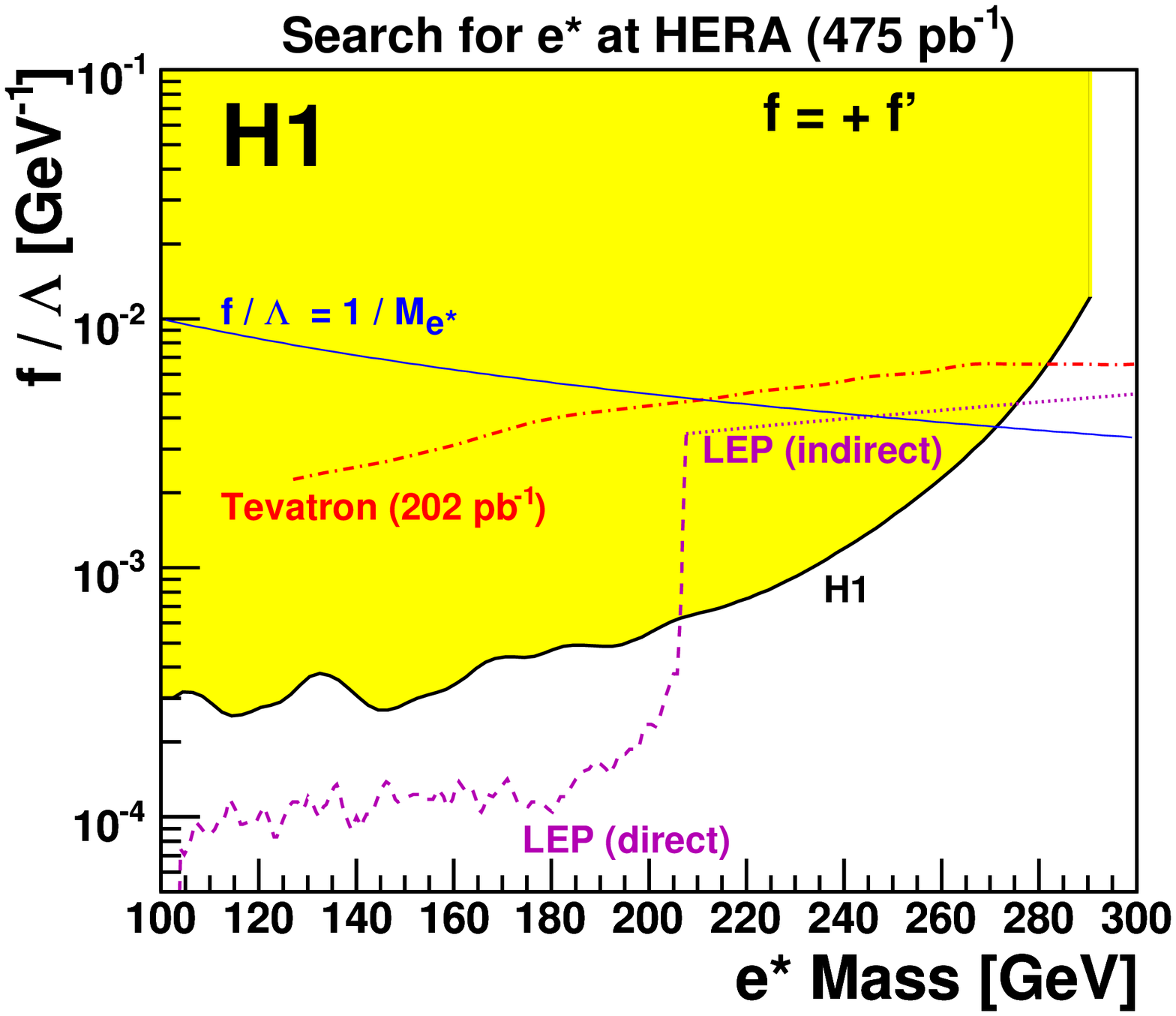}
\includegraphics[width=0.41\columnwidth]{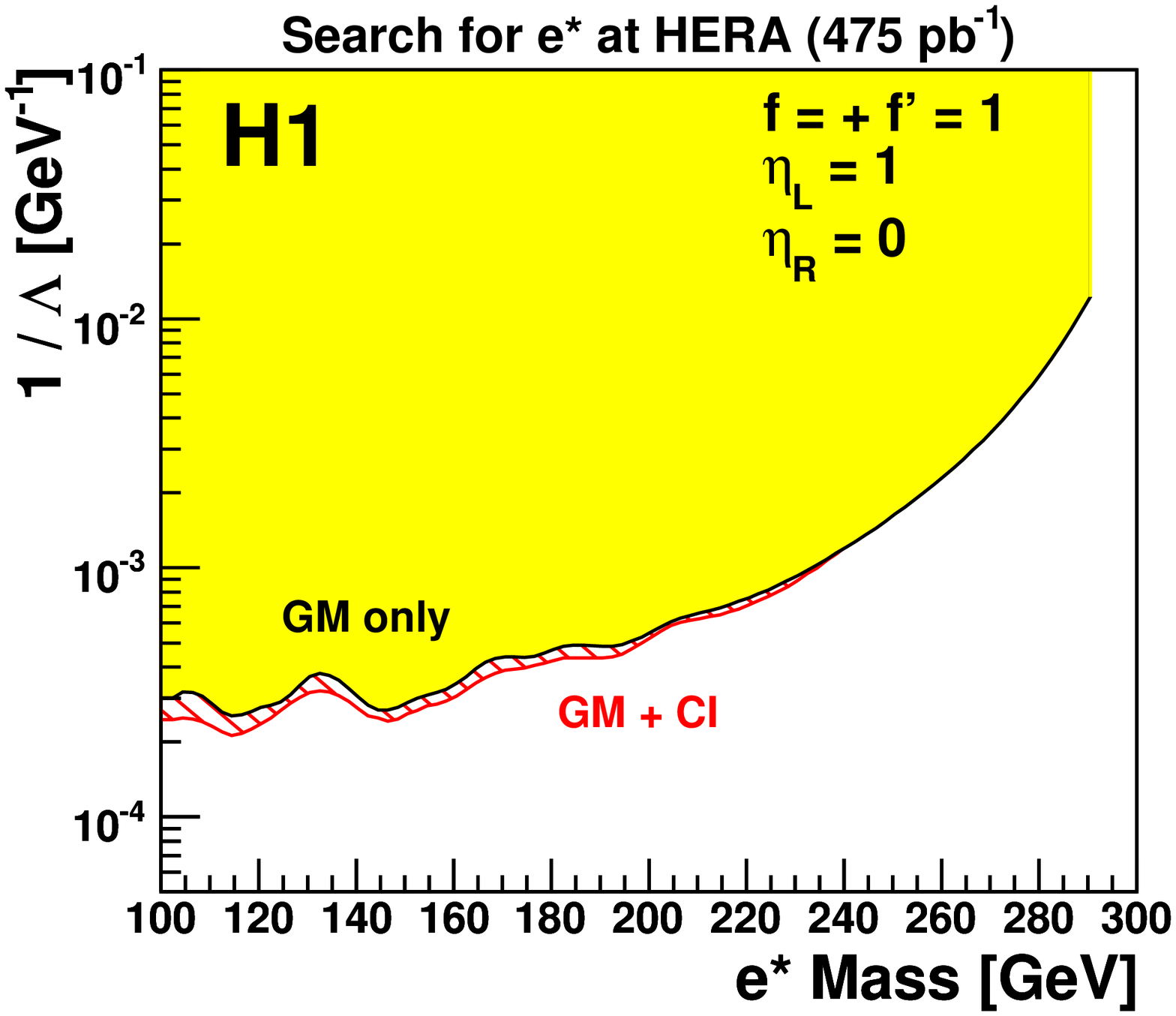}
\caption{ Left: H1 exclusion limits at $95\%$~C.L. on $f/\Lambda$ as a function of the mass of the excited electron
  considering GM interactions only, with the assumption $f = +f'$. The excluded domain is represented by the shaded area.
  Right: H1 exclusion limits at $95$\% CL on the inverse of the compositeness scale $1/\Lambda$ as a function of the mass
  of the excited electron. The hatched area corresponds to the additional
  domain excluded if GM and CI $e^*$ production are considered together.}
\label{fig:estar}
\end{figure*}

The excited electron decays ${e}^{*} \rightarrow e\gamma$,
${e}^{*} \rightarrow eZ$ and ${e}^{*} \rightarrow \nu W$ with subsequent
hadronic or leptonic decays of the $W$ and $Z$ bosons are examined at
HERA in a search for excited electrons by H1~\cite{h1excitedelec}.
In this search, which uses the complete H1 $e^{\pm}p$ data sample, each
decay channel is investigated and no indication of a signal is observed.
New limits on the production cross section of excited electrons are
obtained within a GM model~\cite{excitedGM}, where an upper limit on
$f/\Lambda$ as a function of the excited electron mass is
established for the specific relation $f = +f'$ between the couplings.
Assuming $f = + f'$ and $f/\Lambda=1/M_{e^*}$ excited electrons with a
mass lower than $272$~GeV are excluded at $95\%$~C.L, as shown in
figure~\ref{fig:estar}~(left).
Also shown are direct~\cite{lepdirectexcitedelec} and
indirect~\cite{lepindirectexcitedelec} exclusion limits obtained at LEP,
as well as a result from the Tevatron~\cite{cdfexcitedelec}.
For the first time in $ep$ collisions, gauge and four--fermion contact
interactions are also considered together for $e^*$ production and
decays, assuming $f = +f' = 1$.
The CI term improves the limit on $1/\Lambda$ only slightly,
demonstrating that the GM mechanism is dominant for excited electron
processes at HERA, as illustrated in figure~\ref{fig:estar}~(right).

\begin{figure*}[t]
\centering
\includegraphics[width=0.41\columnwidth]{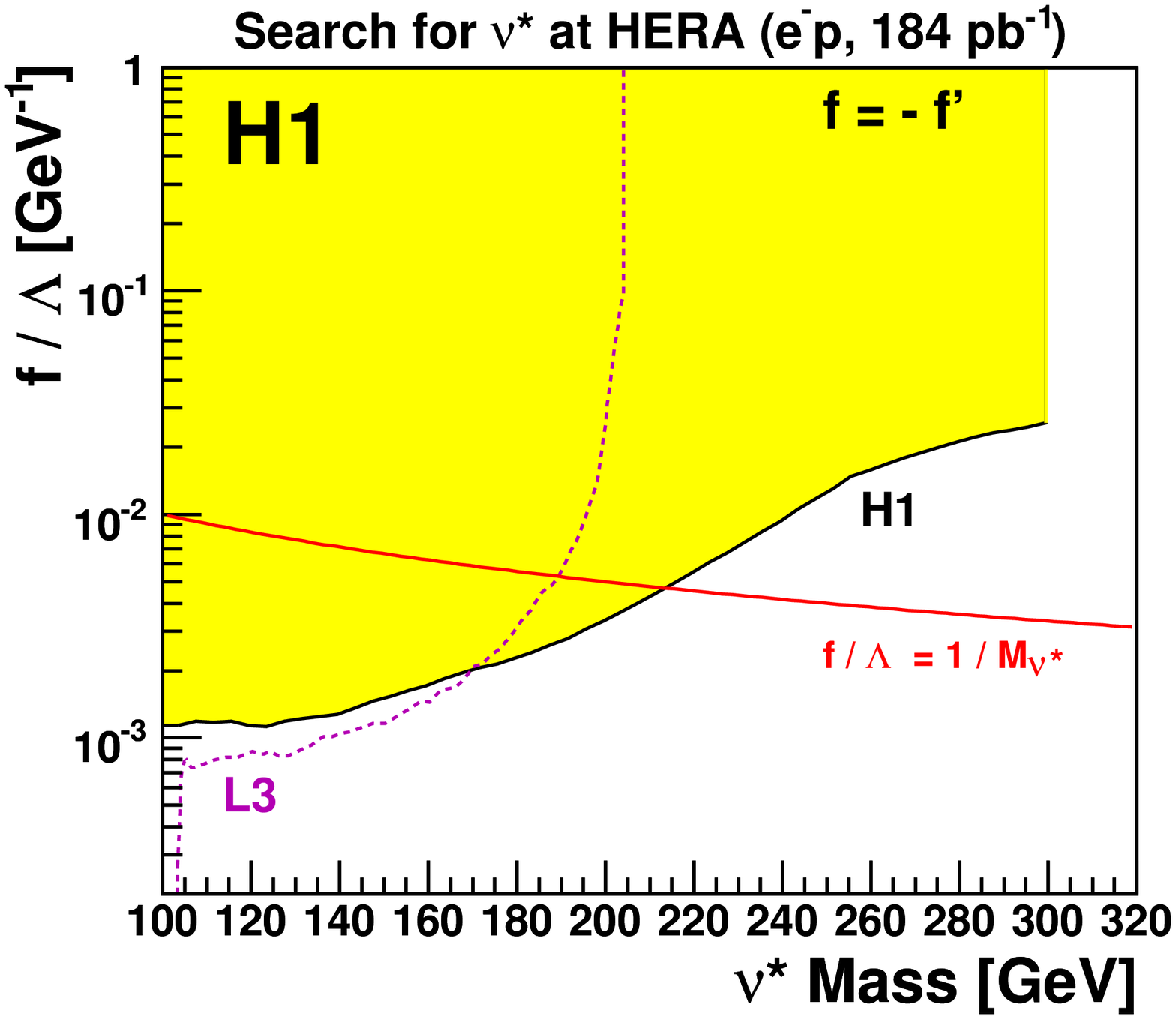}
\includegraphics[width=0.41\columnwidth]{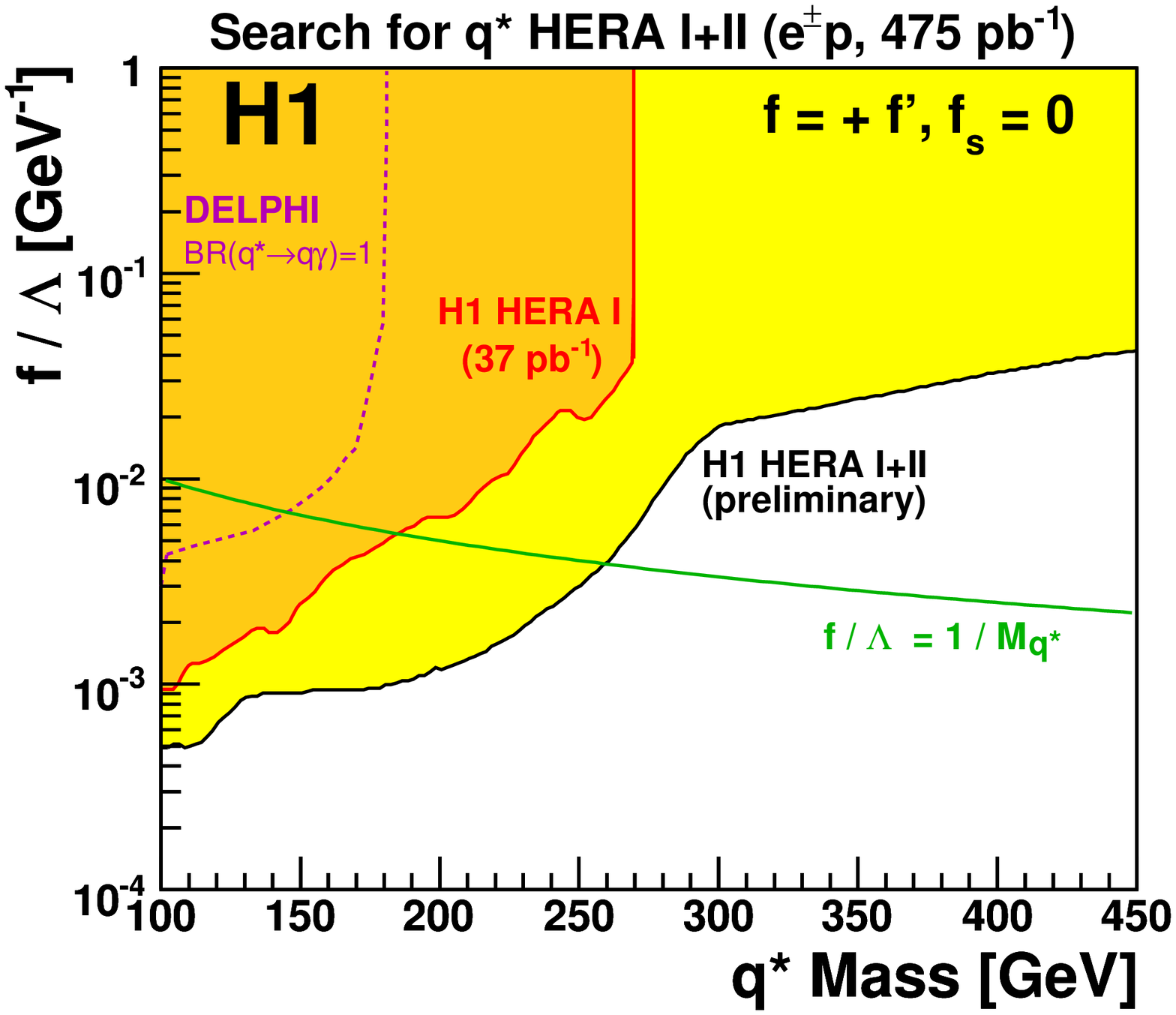}
\caption{Left: H1 exclusion limits at $95\%$ C.L. on the coupling $f/\Lambda$ as a function of the mass of
  the excited neutrino with the assumption $f = -f'$. The excluded domain is represented by the shaded area.
  Right: H1 exclusion limits at $95\%$ C.L. on the coupling $f/\Lambda$ as a function of the mass of
  the excited quark with the assumptions $f = +f'$, $f_{s}=0$. The excluded domain is represented by the shaded area.
  The improvement with respect to previous HERA~I H1 results~\cite{h1excitedq00} is also visible.}
\label{fig:nuqstar}
\end{figure*}

A search for the production of excited neutrinos is performed by H1,
using the full $e^{-}p$ data sample, integrated luminosity
$184$~pb$^{-1}$~\cite{h1excitednu}.
Due to the helicity dependence of the weak interaction and given the
valence quark composition and density distribution of the proton, the
$\nu^{*}$ production cross section is predicted to be much larger
for $e^-p$ collisions than for $e^+p$.
The excited neutrino decay channels ${\nu}^{*} {\rightarrow}
{\nu}{\gamma}$,  ${\nu}^{*} {\rightarrow} {\nu}{Z}$ and ${\nu}^{*}
{\rightarrow} {e}{W}$ with subsequent hadronic or leptonic decays of
the $W$ and $Z$ bosons are considered and no indication of a $\nu^*$
signal is found.
Upper limits on the coupling $f/\Lambda$ as a function of the excited
neutrino mass are established within the GM model for specific relations
between the couplings.
Assuming $f = - f'$ and $f/\Lambda=1/M_{\nu^*}$, excited neutrinos
with a mass lower than $213$~GeV are excluded at $95$\%~C.L., as
shown in figure~\ref{fig:nuqstar}~(left).
The best limit obtained at LEP is also indicated~\cite{lepexcitednu}.
The excluded region is greatly extended with respect to previous results,
based on only $15$~pb$^{-1}$ of $e^{-}p$ data~\cite{h1excitednu01}, and demonstrates
the unique sensitivity of HERA to excited neutrinos with masses beyond the LEP reach.

Finally, a search for excited quarks is performed using the full H1 $e^{\pm}p$
data sample~\cite{h1excitedq}.
The electroweak  decays of excited quarks $q^{*} \rightarrow q\gamma$,
$q^{*} \rightarrow qZ$ and $q^{*} \rightarrow qW$ with subsequent hadronic or
leptonic decays of the W and Z bosons are considered.
No evidence for excited quark production is found.
Mass dependent exclusion limits on the ratio $f/\lambda$ are derived within the
GM model, as illustrated in figure~\ref{fig:nuqstar}~(right).
Assuming $f = +f'$, no strong interactions $f_{s}=0$ and $f/\Lambda=1/M_{q^*}$, excited quarks with a
mass lower than $259$~GeV are excluded at $95\%$~C.L.
The best limit obtained at LEP also indicated~\cite{lepexcitedq}, assuming that
the branching ratio $BR$($q^{*} \rightarrow q\gamma$) = $1$.
These limits extend the excluded region compared to previous H1 excited
quark searches~\cite{h1excitedq00}.

\end{document}